\documentstyle[12pt]{article}

\begin{document}
\title{Correlations between black holes\\ formed in cosmic string
breaking}
\author{R. Emparan \thanks{wmbemgar@lg.ehu.es} \\
Departamento de F{\'\i}sica de la Materia Condensada\\ Universidad del
Pa{\'\i}s Vasco, Apdo. 644\\ 48080 Bilbao, Spain}
\date{}
\maketitle %
\setcounter{page}{0}
\pagestyle{empty}
\thispagestyle{empty}
\vskip 1.0in
\begin{abstract}
An analysis of cosmic string breaking with the formation of black holes
attached to the ends reveals a remarkable feature: the black holes can
be correlated or uncorrelated. We find that, as a consequence, the
number-of-states enhancement factor in the action governing
the formation of uncorrelated black holes is twice the one for a
correlated pair. We argue that when an uncorrelated pair forms at the
ends of the string, the physics involved is more analogous to thermal
nucleation than to particle-antiparticle creation. Also, we analyze the
process of intercommuting strings induced by black hole annihilation
and merging. Finally, we discuss the consequences for grand unified
strings. The process whereby uncorrelated black holes are formed yields
a rate which significantly improves over those previously considered,
but still not enough to modify string cosmology.

\end{abstract}
\vfill

\begin{flushleft} EHU-FT/9511\\
gr-qc/9507002\\ June 1995
\end{flushleft}

\newpage\pagestyle{plain} %
\section{Introduction}

Recent investigations on pair production of black holes have yielded new
evidence for the conjecture that black holes have a number of internal
states given by the exponential of one quarter of the area of the event
horizon. Processes where black holes are spontaneously created in
pairs show an enhancement of the probability relative to formation of
e.g., monopoles, although we are still far from having a fully
satisfactory microscopic explanation of the underlying degrees of
freedom.

Black hole pair creation has been investigated in several
scenarios: Schwinger-like production in Maxwell fields
\cite{gs,hhr,jdb,mi}, inflation-driven creation in de Sitter space
\cite{mr} and, very recently, the breaking of a cosmic string with
black holes forming at its ends \cite{break,ehkt,gh}. All these are
quantum tunneling processes mediated by gravitational instantons. In
fact, since these seem to provide a consistent picture of physically
reasonable processes we may take it as an indication that the Euclidean
approach to semiclassical quantum gravity is a meaningful one and that
topology change should be taken into account.

The instantons describing black hole pair production have been mainly
based on different variations of the Euclidean section of the
$C$-metric \cite{kin}. The Lorentzian section of this space describes
two black holes uniformly accelerating away from each other. This
hyperbolic trajectory turns into a circular orbit when continuing to
Euclidean time. One can then proceed in analogy with the
calculation of the Schwinger effect in Ref.\,\cite{aam} and construct
an instanton for computing the rate for pair production in semiclassical
approximation. A nice feature of the general relativistic description
of the process is that the $C$-metric signals by itself the need of a
force to accelerate the black holes, in the form of conical
singularities along the axis joining the black holes. The different
pair production processes mentioned above
correspond to different ways of dealing with these singularities.

In this paper our interest will be focused on situations where the
force is provided by a cosmic string. The process by which a string
breaks by forming a pair of black holes attached to its ends has also
attracted recent attention because of its possible relevance to cosmic
string cosmology. Although vortex solutions are not easily handled
analytically, recent investigations on Nielsen-Olesen vortices piercing
a black hole \cite{anaetal} or ending at it \cite{gh} have shown that
one can consistently model a thin gravitating physical vortex by a
conical deficit in the spacetime geometry. Moreover, it has been argued
that it is possible for a topologically stable string to end at a black
hole \cite{anaetal,ehkt}. Then the $C$-metric can be interpreted as
providing a picture of a pair of black holes formed by the breaking of
a cosmic string and which are subsequently being pulled apart by the
string tension.

One could also consider a seemingly similar process, in which the
cosmic string breaks in such a way that the string tension is exactly
balanced by the gravitational attraction of the black holes. This black
hole--cosmic string configuration has been considered in
Ref.\,\cite{afv} and very recently in Ref.\,\cite{gh}. However, as we
will see below, the similarities between the two processes are rather
superficial and, in fact, differences between them will be more
significant. We will find a main difference in that the created black
holes are correlated in one case and uncorrelated in the other. Also,
the mechanisms involved in string breaking are utterly different in
each process. We will also discuss how black hole
annihilation or merging provides a mechanism for intercommuting
strings and, finally, we will analyze the relevance of the
different breaking processes in the context of grand unified
strings. The idealization will always be made that a thin cosmic string
is well approximated by a conical singularity.

\section{Breaking the string}

Let us first consider summarily the breaking of a cosmic string as
described by the $C$-metric. The leading contribution to the instanton
action governing the semiclassical rate for string breaking,
$\exp(-I)$, was computed in Ref.\,\cite{break} to be
\begin{equation}\label{leading}
I\simeq {\pi m^2\over \mu}\,,
\end{equation}
where $m$ is the black hole mass and $\mu$ the string tension. This
result is the same as the one for breaking the string
forming monopoles at its ends \cite{vil}. To find out
whether substitution of monopoles with black holes results in an
enhanced rate, we will expand to next-to-leading order the exact
result in Ref.\,\cite{break}. For gravitationally corrected
monopoles the action is the same as for creation of extremal black
holes, since the topology of the instantons is the same in both cases.
We find
\begin{equation}\label{monop}
I_{\rm mon}={\pi m^2\over \mu} - \pi m^2+\cdots
\end{equation}
For nearly extremal black holes (the wormhole configuration) the
result is
\begin{equation}\label{wormh}
I_{\rm non-ext}={\pi m^2\over \mu} - 2\pi m^2+\cdots
\end{equation}
The difference between both is, as expected \cite{hhr,jdb}, the factor
$\pi m^2=A_{\rm bh}/4$. Therefore, breaking by non-extremal black holes
is enhanced by a factor of $\exp S_{\rm bh}$.

Let us now consider the process of static breaking of a cosmic string
\footnote{We will refer to this process as `static breaking'; however,
it should be noted that the $C$-metric, although describing
accelerating objects, is also static in a way similar to Rindler
space.}. The metric describing it belongs to a family of axisymmetric
multi-Schwarzschild configurations which are conveniently written in
Weyl's canonical coordinates \cite{ik}
\begin{equation}\label{weyl}
ds^2=-e^{2\psi}dt^2+e^{-2\psi}[e^{2\gamma}(d\rho^2+dz^2)+\rho^2
d\phi^2] \,,
\end{equation}
with
\begin{equation}\label{psi}
\psi={1\over 2}\sum_{i=1}^{N}\ln{r_{i}^{(+)} +r_{i}^{(-)} -2M_i\over
r_{i}^{(+)} +r_{i}^{(-)} +2M_i} \,,
\end{equation}
and
\begin{equation}\label{gamma}
\gamma={1\over 4}\sum_{i,j=1}^{N}\ln {[r_{i}^{(+)} r_{j}^{(-)}
+l_{i}^{(+)} l_{j}^{(-)} +\rho^2]\over [r_{i}^{(-)} r_{j}^{(-)}
+l_{i}^{(-)} l_{j}^{(-)} +\rho^2]} {[r_{i}^{(-)} r_{j}^{(+)}
+l_{i}^{(-)} l_{j}^{(+)} +\rho^2]\over [r_{i}^{(+)} r_{j}^{(+)}
+l_{i}^{(+)} l_{j}^{(+)} +\rho^2]}+\gamma_0 \,.
\end{equation}
Here we have defined
\begin{eqnarray}\label{rpis}
{r^{(\pm)}_i}^2={l^{(\pm)}_i}^2+\rho^2 \,,\nonumber\\
{l^{(\pm)}_i}=z-z_i\pm M_i \,.
\end{eqnarray}
This metric describes $N$ black holes of masses $M_i$, with event
horizons at $\rho=0$, $|z-z_i|\leq M_i$. In general, the metric
possesses conical singularities at the intervals between black holes,
$z_i+M_i <z< z_{i+1}-M_{i+1}$, $\rho=0$. If $\phi$ has periodicity
$2\pi$, then for a small circumference centered at these points
\begin{equation}\label{consin}
{{\rm
circumference}\over {\rm radius}}\bigg|_{{\rm rad}\rightarrow 0}=2\pi
e^{-\gamma}|_{\rho=0}= 2\pi {(z_{i+1}-z_i)^2 -(M_i-M_{i+1})^2\over
(z_{i+1}-z_i)^2 -(M_i+M_{i+1})^2}e^{-\gamma_0}\,,
\end{equation}
while at the axis outside the
black holes ($z<z_1-M_1$ or $z>z_N+M_N$) we have a
conical deficit $2\pi[1-\exp(-\gamma_0)]$. It is therefore impossible
in general to adjust $\gamma_0$ to cancel the conical singularities
everywhere. The physical reason for this is clear: we need a force to
balance the gravitational attraction between the black holes. We can
choose $\gamma_0$ so that we have a conical deficit running from the
outermost black holes to infinity, and smaller (or zero) conical
deficits inbetween the black holes. The physical configuration
described in this way is that of a thin cosmic string split or
`thinned' by a set of black holes.

The most interesting cases correspond to $N=1,2$, also considered in
Ref.\,\cite{afv}.
For $N=1$ and $\gamma_0\neq 0$, we have a black hole of gravitational
mass $M$ and internal energy (ADM mass)  $Me^{-\gamma_0}$ pierced by a
cosmic string of energy density $\mu=(1-e^{-\gamma_0})/4$. For $N=2$
and taking $M_1=M_2=M$, the choice
\begin{equation}\label{gamstr}
\gamma_0=\ln{(\Delta z)^2\over (\Delta z)^2-4M^2}
\end{equation}
($\Delta
z\equiv |z_2-z_1|$),
results in the geometry of a thin cosmic string broken by a pair of
static black holes separated a (coordinate) distance
\begin{equation}\label{deltaz}
\Delta z=M/\sqrt{\mu}\,.
\end{equation}
This equation can also be read as the balance between the string tension
and the Newtonian gravitational attraction. For the thin string limit
to be consistent we must have $M>\mu^{-1/2}$ \cite{anaetal}. Also,
for the semiclassical calculations to be reliable we will need
$M\gg M_{\rm Pl}$ and small $\mu$. One deduces from here that the black
holes must be far from each other, $\Delta z>\mu^{-1}$.

We will focus on this solution, and compute its Euclidean action
to obtain the probability for static string breaking. First we continue
$t\rightarrow i\tau$. To avoid
conical singularities at the horizons ($\exp \psi=0$)
the Euclidean time must have period $\beta=8\pi M$. This amounts to
requiring the black holes to be in a thermal
bath so that they are in thermal equilibrium with their
surroundings (this condition also requires equality of the black hole
masses). In contrast, thermal equilibrium in the situation described
by the $C$-metric is achieved by equating the black
hole and acceleration temperatures and, to obtain this, one must
introduce a charge to lower the black hole temperature to a value close
to extremality \footnote{It has been argued \cite{dgh} that physical
vortices can dress the horizons and relax the periodicity requirements.
This would allow to consider, e.g., an uncharged $C$-metric \cite{gh}.
However, for simplicity we will continue to keep the usual conditions
for Euclidean time periodicity. Our conclusions will not be affected
in any essential way by this.}.

Following
Ref.\,\cite{hh}, we compute the Euclidean action as
\begin{equation}\label{eucact}
I_E=\beta H-{1\over 4} A_{\rm hor} \,.
\end{equation}
The action must be defined relative to a
reference background, which we take to be thermal flat space
containing a thin cosmic string of tension $\mu$, and at the same
temperature. The Hamiltonian $H$ contains two terms: one is
proportional to the Hamiltonian constraint of General Relativity and
vanishes for exact solutions. The other is a boundary term which is
easily computed. We find
\begin{equation}\label{ham}
\beta H=2\beta M(1-4\mu)\,.
\end{equation}
The area term in Eq.\,(\ref{eucact}) appears because surfaces of
constant $\tau$ intersect at the horizons. It contains the
difference between the area of all the horizons with respect to the
reference background. In our case, two black hole horizons of the same
area are present, while the reference metric contains none. For each of
these horizons we find
\begin{equation}\label{area}
A_{\rm bh}=\int_0^{2\pi} d\phi\int_{z_2-M}^{z_2+M} dz
\sqrt{|g_{zz}g_{\phi\phi}|}_{\rho=0}=2\beta M {\Delta z\over \Delta
z-2M} \,.
\end{equation}
Then
\begin{equation}\label{finact}
I_E=\beta M\left(1-{2\sqrt{\mu}\over 1-2\sqrt{\mu}}-8\mu\right)
\end{equation}
and, for small $\mu$, we will keep only the leading term $I_E\simeq
\beta M$.

By slicing in half the Euclidean geometry and gluing it to a constant
Lorentzian time section we obtain an instanton describing the quantum
tunneling from the initial string to a string split by a pair of black
holes. The leading semiclassical value for the rate of this process is
given by $\exp (-I_E)$.

We want to compare this result to the process where the string breaks
by forming at its ends particles without horizons. In contrast to the
$C$-metric process, now we can not consider extremal black holes: the
modification of Eqs.\,(\ref{psi},\ref{gamma}) to include magnetic
charge, i.e., the axisymmetric multi-Reissner-Nordstrom metric, describes a
pair of {\it equally charged} objects whose repulsion tends to balance
the gravitational attraction, doing so completely in the extremal case;
conical singularities and therefore cosmic strings can be avoided in the
latter case. It is also clear that we can not use this form of the
metric if magnetic charges are to be conserved.

On the other hand, if we wanted the string to break into monopoles, it
should be topologically unstable. In any case, string stability will not
be an issue for the arguments in this section, which concern the
properties of black holes and not those of strings. In the next section
we will discuss some aspects of the breaking of real cosmic strings,
but for the moment we will consider that the string can break into two
neutral massive particles without horizons, separated a distance much
greater than their radius. The rate for this process will be given by
Eq.\,(\ref{eucact}) without the area terms, i.e., $\sim\exp(-2\beta M)$.

Therefore we find that the probability for string breaking by black
holes is enhanced by a factor $\exp (2S_{\rm bh})$, where $S_{\rm bh}=
4\pi M^2$ is the entropy of a single black hole. Instead, the
enhancement factor for the $C$-metric is only $\exp S_{\rm bh}$. This
is a clear indication that string breaking by accelerating black holes
and by static ones are markedly different processes. Actually, whereas
the $C$-metric produces a black hole/anti-black hole pair, the static
breaking does not. The Euclidean section of the $C$-metric describes
the circular trajectory of a single black hole in Euclidean time. When
one slices the Euclidean geometry at $\tau={\tau_0}$ and at the
corresponding antipodal value $\tau={\tau_0}+\beta/2$, the spatial
section contains two oppositely charged black holes. In this case we
expect the internal state of one of the members of the pair to be
correlated with the internal state of the other, and therefore we
expect to obtain just one factor $\exp S_{\rm bh}$. Mathematically, this
happens because the Euclidean $C$-metric contains only one black hole
`bolt' (the `acceleration bolt' is dealt with by taking the adequate
reference metric).

In contrast, the Euclidean multi-black hole metric contains $N$
`bolts', each of which contributes a factor $-A_{\rm bh}/4$ to the
action. In this case it seems more proper to regard the formation of two
black holes at the ends of the string not as the creation of a
particle-antiparticle pair, but rather as a process of {\it thermal
nucleation of black holes in a heat bath}, like the one studied in
Ref.\,\cite{gpy}. The Euclidean topology of the static two-black hole
instanton is different from the one required for pair creation (i.e.,
$S^2\times S^2-\{{\rm pt}\}$, with spatial sections $S^2\times
S^1-\{{\rm pt}\}$, for pair creation of non-extremal black holes in a
non-compact space). In the instanton for the static process we have
two different `bolts' corresponding to different horizons and therefore
we do not expect any correlation between the internal states of the
black holes. This explains the enhancement factor in this case. Each
time a black hole is added to the configuration, the leading
semiclassical probability is multiplied by a factor $\sim \exp(-\beta
M+ S_{\rm bh})$.

This interpretation is consistent with what we have said about the
charge of black holes in each process. The $C$-metric instanton
requires the black holes to have opposite charges. The static instanton
would yield, if modified to include charge, two black holes with equal
charges of the same sign, clearly something not expected in
particle-antiparticle creation.

Differences between the two processes not only concern the correlations
between the internal states of the black holes. Actually, the action
(\ref{finact}) is completely dominated by the thermal nucleation
factor. In contrast, the $C$-metric instanton is dominated
by the string action: the leading term in Eq.\,(\ref{leading}) can be
obtained by computing the Nambu-Goto action of the string that is
`eaten' when the string breaks apart (see Ref.\,\cite{mi} for a
similar calculation), and the number-of-states enhancement factor
appears only at next-to-leading order. Therefore, in the static
process the string does not break because of its tension, but rather
because black holes nucleate on it. One could say that in the
accelerating process the black hole pair is created by the breaking of
the string, whereas in the static situation it is the formation of the
black holes what causes the string to break.

What happens to a string swallowed by a black hole? Amusingly, the
string does not actually break. For the $C$-metric process (involving
non-extremal black holes) the string passes through the wormhole
connecting the black holes \cite{ehkt}. Apparently we may run
into trouble in the static process, since we have argued that there is
not such a wormhole. However, no real problem is posed by this. For a
single Schwarzschild solution there are two causally disconnected
asymptotic regions, and the string can pass through the Einstein-Rosen
bridge from one region to the other \cite{anaetal}. In the two-black
hole configuration that we have been considering, the black
holes connect to another asymptotic region, identical to the
initial one (it is easily seen that both black holes must connect the
same asymptotic regions: since the global geometry is static, the
string tension in each asymptotic region must be balanced by the
gravitational attraction of the other black hole). Then, the
string can disappear and reappear in them without needing to connect
the black holes. In this sense, we can not break the string, but only
`hide away' some part of it. If the string disappears down a
non-extremal black hole, it must reappear through another one, either
in the same asymptotic region or in a different one. In contrast, if
the black hole is extremal, the string disappears down its infinite
throat. Finally, when a string is created joining a pair of
non-extremal black holes as in Ref.\,
\cite{mi}, we can think of it as a closed string.

Finally, we point out that the static configuration is unstable. The
reason is that black holes have negative specific heat, and
therefore a small fluctuation would cause them to grow or to
evaporate. The balance between string tension and gravitational
attraction would then be destroyed, and the black holes would
either approach and merge or accelerate away. A small acceleration
temperature would then appear, but it is unlikely that it may
compensate for the evaporation.

\section{Black hole annihilation and intercommuting strings}

We would like to address now processes of annihilation or merging
of black holes created in one of the two ways presented here. Black
hole annihilation has been analyzed in the context of pair creation
processes in Ref.\,\cite{hhr}.

The same instanton that describes pair creation can,
when reversed in time, describe pair annihilation, yielding an equal
rate for the process. It is clear that we can describe in this way the
annihilation of two black holes previously created as a
particle-antiparticle pair. However, as it has been argued in
\cite{hhr}, black hole annihilation between members of different
pairs should also be possible, with a probability to happen of order
$\exp(- S_{\rm bh})$. This crossed annihilation of two pairs
would contain just one black hole loop, instead of the two loops
involved in independent creation and annihilation of the two pairs.
In general, the action for these processes of creation or annihilation
will contain one factor of black hole entropy per each black hole loop.

It is interesting to find that when the black holes are
created at the endpoints of the string, annihilation between members of
different pairs results in a process of intersecting and
intercommuting strings, analyzed in \cite{vil}. Conservation laws
require that these processes of crossed creation and annihilation take
place only between oppositely charged black holes and with zero total
magnetic flux.
Using the instanton methods that we have been describing, one can
compute the probability that two intersecting strings intercommute by
this process. The previous argument implies that intersecting and
intercommuting should be suppressed relative to non-intercommuting by a
factor $\exp(- S_{\rm bh})$.

Instead, we could consider a process whereby the strings intercommute
without annihilating the black holes. In this case, black holes would
merge into a bigger one, and we would end with a
pair of strings each threading a black hole. The probabilities for
intercommuting and non-intercommuting in this case would be of the same
order. However, intercommuting by this process would still be very low
because of the need to first break the string by quantum tunneling,
and it is generally assumed that intercommuting takes place by
a classical mechanism which makes
the probability to be of the order of unity.

If we consider now black holes thermally nucleated at the string ends,
reversed instantons would correspond to thermal fluctuations in
which black holes disappear, not necessarily in pairs. In this case,
once the strings are broken their ends do not know to which other end
they were previously joined (as long as the string tensions and black
hole parameters are the same). We could
still consider direct annihilation with a probability of order
$\exp(-S_{\rm bh})$ but, again, the dominant process for
string intercommuting mediated by black holes would involve merging.

\section{Strings and black holes at GUT scale}

If a string is present in hot flat space, it is more probable
for a black hole to nucleate on it than away from it. The reason is
that, for a given temperature, $T=(8\pi M)^{-1}$, the energy required to
nucleate a black hole piercing a string is
\begin{equation}\label{eint}
E=M_{\rm int}=M(1-4\mu)<M\,,
\end{equation}
(this can
be calculated easily using the expression for the energy in
Ref.\,\cite{hh}; see also Refs.\,\cite{afv,anaetal}). The action in
this case is $I=\beta M_{\rm int}/2$, and there is an enhancement over
nucleation in flat space given by $\delta I=2\beta M\mu$.

The temperature at which grand unified strings are expected to form is
$T_{\rm GUT}\sim v\sim 10^{16} {\rm GeV}$, where $v$ is the Higgs
vacuum expectation value corresponding to string
formation. The tension of the strings would then be $\mu\sim v^2\sim
10^{-6} M_{\rm Pl}$. At this temperature, we expect to find nucleation
of black holes of mass $M\simeq 1/(8\pi T_{\rm GUT})\sim 1/(8\pi
\sqrt{\mu})$.
However, for the thin string limit to hold we should let the
temperature lower down at least one order of magnitude. We will
consider $M{\;\lower3pt\hbox{$\buildrel > \over \sim$}\;} \mu^{-1/2}$.
The mass of these black holes would be $M\sim 10^3 M_{\rm Pl}$, still
large enough for the semiclassical approximation to be reliable. In
this case,
\begin{equation}\label{gutact}
I\simeq {1\over 2} \beta M \geq O(\mu^{-1})\sim 10^6\,,
\end{equation}
which yields a negligible nucleation rate. However, the relative ratio
for nucleation at a string compared to nucleation in flat space is
$\sim\exp(16\pi M^2\mu){\;\lower3pt\hbox{$\buildrel > \over
\sim$}\;} 1$. Thus, when strings are present, if black hole
nucleation were to occur, it would mainly take place at a string.

Let us now consider the process of string breaking. In
Refs.\,\cite{break,ehkt} the rate for the process mediated by the
$C$-metric instanton has been calculated using Eq.\,(\ref{leading});
an approximate estimation yields $I>10^{12}$, but this is only a
lower limit, since one expects that the dominant process should
involve breaking by black holes smaller than the string thickness. A
more detailed examination in Ref.\,\cite{gh} shows that it may be
possible to lower the value down to $I\sim 10^7$.

Now we want to find out whether the breaking
induced by thermal black hole nucleation can modify these rates. An
issue to clarify previously is whether it is more probable for
the string to break than to `thin' by nucleating black holes a distance
larger than (\ref{deltaz}), in which case a residual string
tension $\mu_{\rm in}$ would remain inbetween the black holes.

It is not difficult to find that when $\gamma_0$ is selected so as to
match the required string tensions, the following relation must hold
\begin{equation}\label{muin}
{M\over \Delta z}=\sqrt{\mu-\mu_{\rm in}\over 1-4\mu_{\rm in}}\,.
\end{equation}
The Hamiltonian is still given by Eq.\,(\ref{ham}), but the black hole
area is modified. The final result for the action is
\begin{eqnarray}\label{fray}
I&=&2\beta M(1-4\mu)-\beta M{1+2M/ \Delta z\over
1-4\mu}\nonumber\\
&=& \beta M \left(1-2\sqrt{\mu}+{\mu_{\rm in}\over
\sqrt{\mu}}+\cdots\right) \,,
\end{eqnarray}
where the small $\mu$ limit has been taken in the last line. It is
evident from here that the action takes its minimum value for
$\mu_{\rm in}=0$.

Then, at finite temperatures and in the presence of cosmic strings, we
could expect to find black hole nucleation at the strings in such a way
that the strings break. The value of the action governing this process
at GUT temperatures is $I{\;\lower3pt\hbox{$\buildrel > \over
\sim$}\;} 10^{6}$,  significantly lower than the naive estimation based
on Eq.\,(\ref{leading}), but still negligible. It is unlikely that the
arguments considered in Ref.\,\cite{gh} to improve
Eq.\,(\ref{leading}), which involve the strong magnetic fields present
near the string core, would be of use in this situation, since, as we
have seen, it is essentially the thermal bath and not the string what
causes black hole nucleation. Therefore, even if the mechanism we have
been considering could improve over those previously studied, breaking
by black holes is not likely to modify in any essential way string
cosmology.

\medskip

\section*{Acknowledgements} Conversations with Ana Ach\'ucarro,
I{\~n}igo Egusquiza and Juan L. Ma{\~n}es are gratefully acknowledged.
This work has been partially supported by a FPI grant from MEC (Spain)
and projects UPV 063.310-EB119-92 and CICYT AEN93-0435.

\end{document}